\documentclass[runningheads]{llncs}
\usepackage{graphicx}
\usepackage[hidelinks]{hyperref}

\usepackage{booktabs}       
\usepackage{makecell}       

\newcommand{\mytilde}{\raise.17ex\hbox{$\scriptstyle\mathtt{\sim}$}}

\begin{document}
\title{Mining Issue Trackers: Concepts and Techniques}
%
%
\author{Lloyd Montgomery,
Clara L{\"u}ders,
Walid Maalej
}
\authorrunning{Montgomery et al.}
%
\institute{University of Hamburg\\Department of Informatics\\Vogt-Kölln-Str. 30, Hamburg 22527, Germany\\
\email{\{lloyd.montgomery,clara.marie.lueders,walid.maalej\}@uni-hamburg.de}}

\maketitle              
\begin{abstract} 
An issue tracker is a software tool used by organisations to interact with users and manage various aspects of the software development lifecycle.
With the rise of agile methodologies, issue trackers have become popular in open and closed-source settings alike.
Internal and external stakeholders report, manage, and discuss ``issues'', which represent different information such as requirements and maintenance tasks.
Issue trackers can quickly become complex ecosystems, with dozens of projects, hundreds of users, thousands of issues, and often millions of issue evolutions.
Finding and understanding the relevant issues for the task at hand and keeping an overview becomes difficult with time.   
Moreover, managing issue workflows for diverse projects becomes more difficult as organisations grow, and more stakeholders get involved.
To help address these difficulties, software and requirements engineering research have suggested automated techniques based on mining issue tracking data.
Given the vast amount of textual data in issue trackers, many of these techniques leverage natural language processing.
This chapter discusses four major use cases for algorithmically analysing issue data to assist stakeholders with the complexity and heterogeneity of information in issue trackers.
The chapter is accompanied by a follow-along demonstration package with JupyterNotebooks.
\end{abstract}


\section{Introduction}

It is rare to find a software organisation today that does not use an issue tracker to support some---if not all---of their processes. 
Users, developers, testers, and managers use issue trackers to report and collect issues such as bug reports.
Subsequently, the issues get discussed, prioritised, assigned as a unit of work, tracked, and eventually resolved, thereby resulting in updated software. 
In addition to bug reports, issue trackers nowadays often collect other issue types including epics, user stories, tasks, and change requests.
This diversity of information types makes these tools indispensable for various software processes including requirements elicitation, bug fixing, product design, user support, and quality assurance.
Issue trackers are so adaptive that, in reality, they can store anything that needs to be tracked and that can fit into the schema of a title, description, and metadata.

\begin{figure}[!ht]
    \centering
    \includegraphics[width=\textwidth]{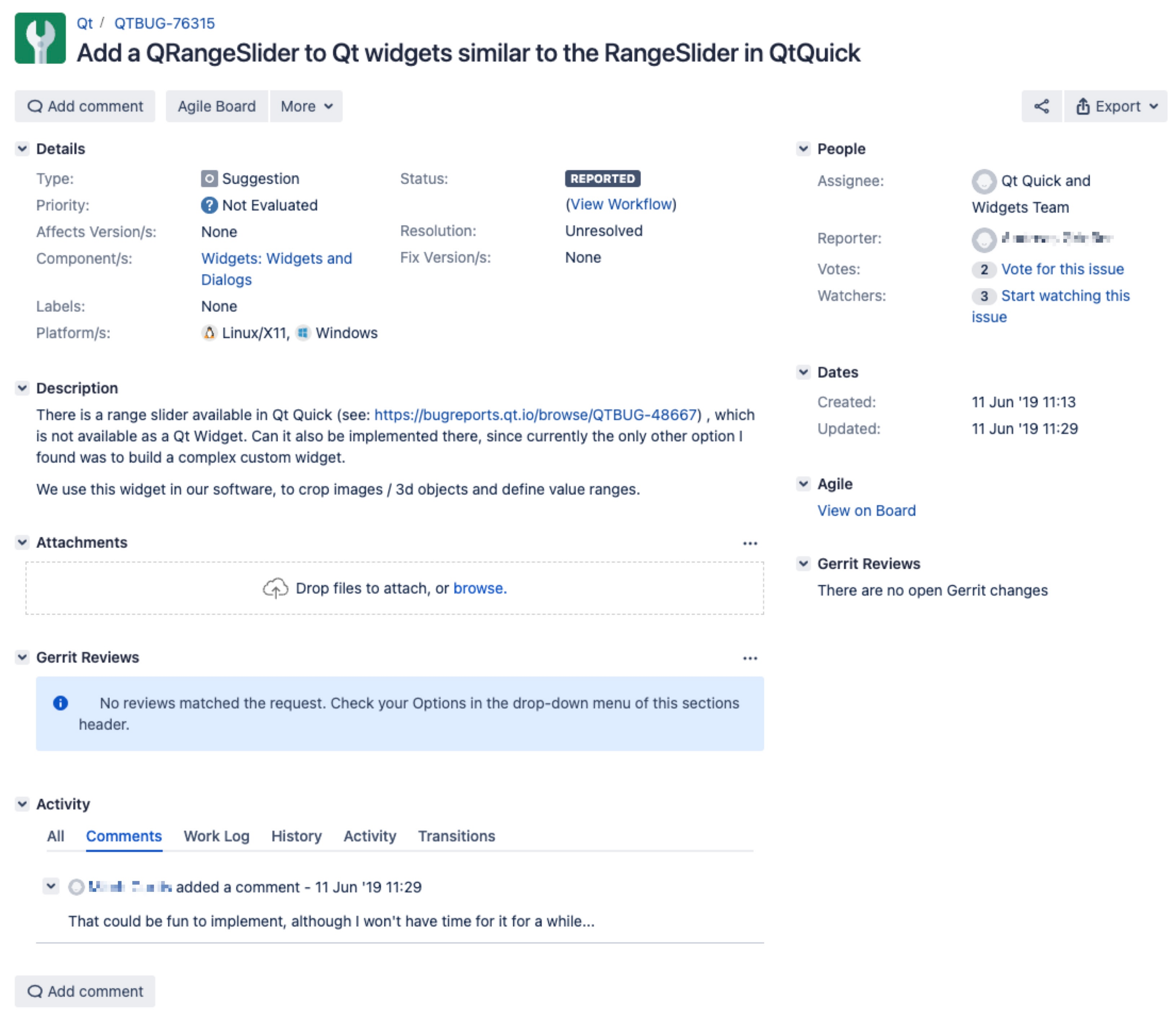}
    \caption{An example of an issue report in Jira.}
    \label{fig:JIRAExample}
\end{figure}

Once filed, issues can contain several supported fields such as issue type, status, comments, severity, priority, project, components, labels, issue links, sub-tasks, creator, reporter, assignee, etc.
What makes issues so powerful is that they are atomic units \textit{separated} from the rest of the project (and can thus be created, viewed, and edited in isolation). When desired, issues can also be \textit{linked to} the project or the corresponding product in different ways.
They can be grouped together under components, share a set of labels with other issues, be directly structured under other issues as sub-issues, or be linked to other issues  denoting various types of dependencies.
Figure~\ref{fig:JIRAExample} shows an example of an issue in the Qt Jira issue tracker: QTBUG-76315.\footnote{\url{https://bugreports.qt.io/browse/QTBUG-76315}}
The title is displayed across the top.
The other fields are scattered around the interface, including the issue ``Type'' (Suggestion), the short description, and the comments at the bottom.

One fundamental defining difference between issues is the chosen value for the ``Issue Type'' field, which can, for example, be ``Bug Report'', ``Epic'', ``User Story'', ``User Support'', etc.
The description for an Epic is to be interpreted differently than the description for a User Story, which is to be interpreted differently than a Bug Report, etc.
In other words, the Issue Type clearly separates issues into different functions for the software development lifecycle and how the issues should be tracked and managed.
Montgomery et al.~\cite{Montgomery_MSR_2022} performed a thematic analysis of unique issue types across various public Jira issue trackers and found that the majority of them can be grouped into four issue categories: 
\begin{itemize}
    \item \textbf{Requirements issues} including epics, user stories, as well as feature requests and change requests.
    \item \textbf{Development issues} document the development work breakdown into assigned tasks, including technical tasks, development tasks, and sub-tasks.
    \item \textbf{Maintenance issues} including reported bugs, defects, incidents, technical debts, or documentation issues.
    \item \textbf{User support issues} representing  dialogues between developers and users, as support requests, problem tickets, IT help requests, or questions.
\end{itemize}

With time, the vast amount of generated and continuously evolving software project data gets reflected in issue trackers and often overwhelms their users~\cite{Baysal_ICSE_2013,Anvik_OOPSLA_2005}.
Issue trackers can accumulate thousands or even millions of reports throughout the lifetime of a project~\cite{Anvik_OOPSLA_2005,Regnell_REFSQ_2008}.
Such overload of issue information might itself impact  productivity and decision-making~\cite{Ho_GROUP_2001,Fucci_ESEM_2018}.
Moreover, the flexibility of issue trackers and the low usage barrier for stakeholders with different backgrounds, skills, and motivations can lead to  redundant~\cite{zhang2023duplicate}, ambiguous~\cite{ferrari2019nlp}, or conflicting issue  data~\cite{Zimmermann_TSE_2010}.
Possible consequences include poor issue report quality, intricate issue triage, delayed issue resolution, and communication difficulties ~\cite{Baysal_FSE_2014,Cavalcanti_JSEP_2014,Zou_TSE_2020,Zhang_SCIS_2015}.

To address the challenges of an overwhelming amount of heterogeneous and often ambiguous issue-related information, Natural Language Processing (NLP) techniques bear a large potential~\cite{zhao2021natural}.
This chapter delivers a detailed overview of issue tracking data, along with algorithmic approaches to support  stakeholders dealing with these challenges.
We first discuss popular issue trackers, their main usage, as well as the structure and semantic of data that is normally maintained therein.
Then, we discuss four use cases for using NLP techniques on issue tracking for the analysis of requirements quality, issue evolution, discussions, as well as link and traceability analysis.
The common enabler is to combine text data with meta and historical data.

\section{Fundamentals of Issue Trackers}
In this section, we discuss the importance of issue tracking for software organisations and compare popular issue trackers and publicly available issue datasets.
We also describe the data normally found in most issue trackers.

\subsection{Roles of Issue Trackers}

Issue trackers are designed to assist stakeholders in communicating, documenting, and collaborating on project-related issues during the entire software development lifecycle.
It is difficult to think about a software organisation or an open-source community without also thinking about their Jira, GitHub, Bugzilla, or Redmine repository.
These tools play an important role from the Requirements Engineering (RE) and Software Engineering (SE) perspective:

\subsubsection{Issue Elicitation and Just-In-Time RE.}
The primary goal of issue trackers is to collect issues as they arise.
Usually, issues can be reported by various stakeholders: both by the core software team including developers, managers, and product owners, as well as by users, customers, and other stakeholders.
From an RE perspective, issue trackers created a direct avenue for just-in-time requirements~\cite{ernst2012case} and crowd-based requirements~\cite{groen2017crowd,Haering:ICSE:2021}.
With properly managed and maintained issue trackers, a constant stream of just-in-time ``requirements'' can emerge from the community: first filed as requests or reports and then discussed, clarified, and eventually accepted as requirements.
Recent approaches also bring user feedback and community data (e.g. such as tweets and product reviews) into the issue trackers \cite{Haering:ICSE:2021}.
In addition, requirements elicited through conventional activities such as interviews and workshops can be added by analysts and product owners, e.g. as epics, user stories, features, or use cases.
Overall, issue trackers support both traditional and fast, just-in-time, crowd-based requirements elicitation and management through features such as ease-of-use reporting, attaching and linking additional information, and open discussions in the comments.

\subsubsection{User Support.}
Issue trackers are sometimes also called ticket systems: denoting user requests being processed.
One reason issue trackers have gained popularity is their ability to directly support users alongside other related development activities.
User support differs from other tasks such as managing bug and security reports in that it primarily involves the questions, concerns, thoughts, and feelings of the users.
This is a central component of customer relationship management~\cite{reinartz2004customer}.
In other words, user support is as concerned with the user's state of mind as with the software issue they are experiencing.
Issue trackers can support this workflow---publicly or privately, and allow the communication, notification, and direct linking to related bug reports, requests, release schedules, etc.
This creates forward traceability (and accountability) from user concerns to organisational action, and backwards traceability (rationale) for related organisational decisions.

\subsubsection{Communication and Collaboration.}
Organisations, particularly open-source communities, use issue trackers as a central place for communication and collaboration~\cite{Bertram_CSCW_2010,Mockus_TOSEM_2002}.
The central communication tools for communities were once mailing lists and forums, where each thread represented a single discussion about a bug, feature, or wider topic.
Now, many of those discussions are held in the comments sections of corresponding issues,  assisted by further features such as labelling, linking, or watching certain issues and their discussions \cite{Arya_ICSE_2019}.
Beyond communication, the workflow management built around issues facilitates collaboration within teams and with external stakeholders. 
For instance, once an issue gets resolved by a developer, a workflow can be configured to notify someone to review, integrate, or conduct manual acceptance test (and activate corresponding issues).
Similarly, once a task gets blocked or certain actions and clarifications are needed, particular discussions or workflows can be triggered, e.g.~based on issue status, priority, or component that has changed.

\subsubsection{Git-Extended Knowledge Repository.}
Modern issue trackers can be considered extended knowledge repositories for software  organisations. 
New members joining the project or the organisation can start familiarising themselves by checking current issues, as well as the overall state of the issue tracker~\cite{Stanik:ICSME:2018}.
In addition to information about the projects, organisation, and people, issue trackers include other integrated information about the systems under development including the code, versions, documentation, and binaries~\cite{Maalej:ASE:2009,maalej2013managing}.
While code repositories using Git or SVN capture most of what is necessary when it comes to versioning the code and software components, it rarely captures or links the other project artefacts such as requirements, bug reports, or a roadmap.
On the other hand, issue trackers enable a rich knowledge repository accessible by different stakeholders due to their low technical barrier.
For instance, code changes resulting from an issue are usually traced to Git commits and the rationale behind them (whether for bug fixing, feature enhancement, or any other reason). Similarly, binaries, release notes, tests and quality artefacts are either managed or linked to issues and projects as well.

\subsubsection{Project and Workflow Management.}
Modern issue trackers usually offer functionality to plan and manage projects, releases, and roadmaps based on the issues. 
Some issue trackers also offer advanced features to implement a Kanban or a Scrum process. 
Since the issues are already in the repository, this is a pragmatic step instead of using a separate project management tool.
Issue trackers have also enabled a transparent and configurable approach to managing stakeholders' workflows through issue statuses and  conversations.
The community involved in (or just interested in) the development of a particular product can, for example, see the workflow of requirements as they are created, discussed, assigned to sprints, worked on, tested, and delivered.
This transparency allows interested stakeholders to directly identify the software version a particular requirement is attached to, and know when and how the outcomes of that requirement will be delivered.
Similarly, stakeholders interested in particular bug reports can see if their problem is already reported, if similar bug reports have similar details, they can check the status and know if it is being addressed or ignored, and they can also see when the finalised fix for that bug is released.

\subsection{Examples of Issue Trackers}

Among advanced issue trackers known for their extensive features such as customisable issue fields, release planning, and project management, notable options include Jira,\footnote{\url{https://www.atlassian.com/software/jira}} Redmine,\footnote{\url{https://redmine.org/}} Mantis,\footnote{\url{https://www.mantisbt.org/}} Bugzilla,\footnote{\url{https://www.bugzilla.org/}} and GitHub.\footnote{\url{https://github.com/}} Table~\ref{tab:ITScomparison} presents a  comparative analysis based on high-level feature coverage. 
These issue trackers manage diverse issue types such as ``Epic'', ``Bug Report'', ``User Story'', and ``Task'', allowing users to categorise and structure their issues effectively. 
Additionally, they enable users to define the scope of issues within the product or specific components while offering a spectrum of statuses.
However, it's worth noting that disparities exist among these platforms. 
Bugzilla and Mantis, for instance, lack support for custom issue types or statuses, constraining their users to the predefined ``Bug'' type. 
In contrast, both Redmine and Jira provide the flexibility of defining custom issue types and statuses, catering to the nuanced requirements of diverse projects.

\begin{table}[t!]
\setlength\tabcolsep{2.5pt}  
\centering
\caption{{Comparison across popular issue trackers~\cite{Raatikainen_TSE_2022}.}}
\label{tab:ITScomparison}
\begin{tabular}{@{}lccccc@{}}
\toprule
{High-level feature} & {JIRA} & {Redmine} & {Mantis} & {Bugzilla} & {GitHub} \\
\midrule
{Custom fields}  & {X} & {X} & {X} & {X} & {X}\\
{Custom dashboards}  & {X} & {X} & {-} & {X} & {X} \\
{Custom workflows}  & {X} & {-} & {X} & {X} & {-}\\
{Issue links} & {X} & {X} & {X} & {X} & {X} \\
{Link typology} & {X} & {X} & {X} & {X*} & {-}\\
{Duplicate management} & {X} & {X} & {X} & {X} & {X} \\
{Release management} & {X} & {-} & {-} & {-} & {X} \\
{Add-ons/plugins} & {X} & {X} & {X} & {X} & {-}\\
{API integration} & {X} & {X} & {X} & {X} & {X}\\
\bottomrule
\multicolumn{6}{l}{\footnotesize{\makecell[l]{
* Except depends-on, blocks, see-also, and duplicate, link types\\are difficult to customise.
}}} \\
\end{tabular}
\end{table}

Karre et al.~\cite{Karre_ARXIV_2017} conducted an in-depth analysis of 31 prominent issue-tracking tools to identify their features and distinctions. The authors classified these 31 OSS issue-tracking tools into four distinct clusters based on various features including API support, test plan integration, customisable workflows, and custom field support.
Jira was not included in this analysis~\cite{Karre_ARXIV_2017}.
Cluster 1 includes tools that offer a wide range of features, such as test plan integration, customisable workflows, and product roadmap planning.
Notable tools in this cluster includes Bugzilla, Mantis, and Redmine.
Cluster 2 consists of simpler yet highly effective tools known for their strong support for code repositories and localisation.
Prominent tools in this cluster includes GitHub and BitBucket.
Cluster 3 excels in providing robust authentication features and standard reporting capabilities.
Noteworthy tools in this cluster includes BugTraq and JTrac.
Cluster 4 focuses on features related to notifications and command-line support, including Trac, among others.

\subsection{Issue Tracking Datasets}

\begin{table}[t]
\setlength\tabcolsep{2.5pt}  
\centering
\newcommand{\turl}[1]{\tiny\url{#1}}
\caption{Issue Tracker Datsets}
\label{table:issue_tracker_datasets}
\scriptsize
\begin{tabular}{@{}lllrl@{}}
\toprule
    Dataset & Year & Tracker & \# Issues & URL \\
\midrule
    Lamkanfi et al.         & 2013 & Bugzilla & 214,908 & \turl{https://github.com/ansymo/msr2013-bug_dataset} \\
    Lazar et al.            & 2014 & Bugzilla & 1,674,985 & \dag \turl{http://www.csis.ysu.edu/~alazar/msr14data} \\
    Zhu et al.              & 2016 & Bugzilla & 774,809 & \turl{https://github.com/jxshin/mzdata} \\
    Rahim et al.            & 2017 & Redmine & 13,820 & \dag \turl{https://github.com/shamsurrahim/RedmineDataset} \\
    Gousios                 & 2013 & GitHub & --- & \turl{https://github.com/ghtorrent} \\
    Kallis et al.           & 2022 & GitHub & 803,417 & --- \\
    Nikeghbal et al.        & 2023 & GitHub & --- & \turl{https://github.com/kargaranamir/girt-data} \\
    Ortu et al.             & 2015 & Jira & 700,000 & \dag \turl{http://openscience.us/repo/social-analysis/socialaspects.html} \\
    Tawosi et al.           & 2022 & Jira & 508,963 & 
    \turl{https://github.com/SOLAR-group/TAWOS} \\
    Montgomery et al.       & 2022 & Jira & 2,686,282 & \turl{https://zenodo.org/doi/10.5281/zenodo.5882881} \\
    Diamantopoulos et al.   & 2023 & Jira & --- & --- \\
\bottomrule
\multicolumn{5}{l}{\makecell[l]{
\scriptsize \dag These URLs were already broken at the time of publishing this chapter.}}
\end{tabular}
\end{table}

Despite the large amount of research in the area of issue trackers, access to large and rich issue tracker datasets remains rather limited. 
The GitHub\footnote{\url{https://github.com/}} issue tracker hosts millions of public repositories.
The status quo for most other issue trackers, however, is private hosting, often locally hosted on institutional servers, accessed through hidden or intranet domains.
While GitHub is a good source of issue  repositories, it also represents only one specific issue tracker, lacking some common features such as custom linking, issue types, workflows, etc.
Over the last decade, researchers have collected and released several issue tracker datasets---most of the time to address a specific research goal.

Lamkanfi et al.~\cite{Lamkanfi_MSR_2013} built a \textbf{Bugzilla} dataset containing only genuine bug reports from Eclipse and Mozilla projects.
The dataset covers the complete lifetime of each bug report, including the updates for each field.
This dataset can be used for severity prediction, studying bug-fixing time, or wrongly assigned components.
Lazar et al.~\cite{Lazar_MSR_2014} built a dataset for duplicate detection that encompasses data from the Bugzilla repositories of Eclipse, Open Office, NetBeans, and Mozilla.
The dataset contains duplicate issues as well as random non-duplicates. It is used for researching duplicate detection models.
Zhu et al.~\cite{Zhu_MSR_2016} also built a dataset from Mozilla's issue tracker.
Their dataset contains issue reports properties, activities, and comments.
They used the dataset for issue assignment, issue fixing time, and developer participation prediction.

Rahim et al.~\cite{Rahman_JTEC_2017} built a dataset from \textbf{Redmine} to examine issue starvation, meaning that issues with lower severity or priority do not get enough resources to be resolved.
Their dataset encompasses issues and issue-report properties.

Gousios~\cite{Gousios_MSR_2013} created the popular and extensive GHTorent dataset of \textbf{GitHub} data, which encompasses various projects, issues, users, commits, and pull-request information.
The dataset can be used for collaboration and network analysis studies.
Kallis et al.~\cite{Kallis_NLBSE_2022} created a GitHub dataset with 803,417 issues reports extracted from 127,595 open-source GitHub projects for a research competition hosted by the Natural Language-based Software Engineering Workshop (NLBSE).
The dataset encompasses issue report information, which was used for issue type classification.
Nikeghbal et al.~\cite{Nikeghbal_MSR_2023} also mined data from GitHub to create an issue dataset that enable examining the usage of issue templates.

Ortu et al.~\cite{Ortu_PROMISE_2015} created a \textbf{Jira} dataset for Apache, Spring, JBoss, and CodeHaus communities to analyse the communication processes.
This dataset contains issue reports and corresponding comments.
Tawosi et al.~\cite{Tawosi_MSR_2022} crawled public Jira repositories that use agile approaches.
They mined more than half a million issues from 44 open-source projects.
This dataset can be used, e.g., for effort estimation, issue prioritisation, or issue assignment.
Montgomery et al.~\cite{Montgomery_MSR_2022} built a large Jira dataset with the issue reports from 16 different organisations and all their projects, resulting in 2.7 million issues, 32 million changes, 9 million comments, and 1 million issue links.
This is a rather general purpose dataset, which can be used for various goals including issue link prediction, discussion, or evolution analysis.
Diamantopoulos et al.~\cite{Diamantopoulos_MSR_2023} mined the Jira repository of Apache in 2023 and used topic modelling to enhance the dataset.

\subsection{Issue Tracking Data}
\label{sec:issue_data}
\begin{figure}[t]
    \centering
    \includegraphics[width=\columnwidth]{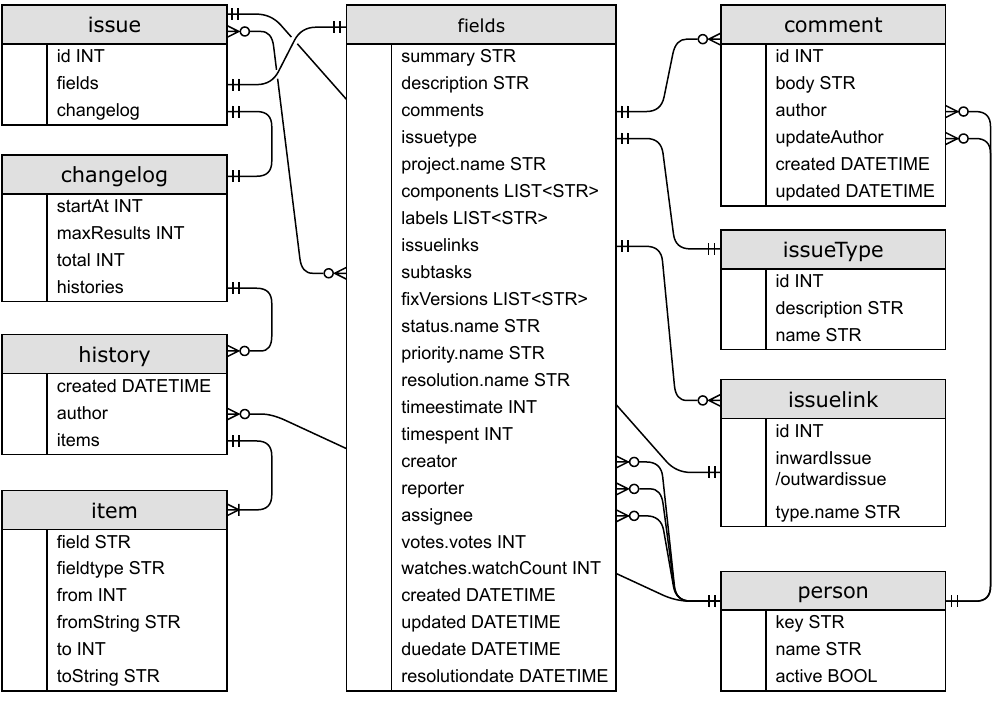}
    \caption{Constructed Jira MongoDB database scheme (from Montgomery et al.~\cite{Montgomery_MSR_2022})}
    \label{fig:database_erd}
\end{figure}

Issue trackers contain an assortment of data from natural language text to structured meta-data.
We summarise five different types of data found in issue trackers and which can be used for different purposes.
Figure~\ref{fig:database_erd} gives an overview of the data usually available in a Jira repository. 
To discuss the data structure, we use Jira as an example, as it is one of the most popular and functionality-rich issue trackers in practice.

\subsubsection{Core Issue Data (Content).}
An issue has two primary fields capturing it core information: Summary, Description. While the data is usually natural text, description can also include formatted text as well as attachments and images. The core issue data is usually available at the issue creation time

The \textit{Summary} (or title) provides a brief overview of the information contained in the issue. 
For example: ``Allow toggling every single projector's ability to cast a shadow separately'' is a summary of STORM-2147 from SecondLife.\footnote{\url{https://jira.secondlife.com/browse/STORM-2147}}
The Summary field alone is often used for tasks such as searching for issues, prioritising issues, assigning issues to people, and managing backlogs.
Therefore, ideally, the Summary should enable a good quick interpretation of the issue overall.
When written effectively, the Summary should touch on all important high-level aspects of the issue and be informative for automated analysis.
In most cases, however, the Summary is not enough for a complete understanding of the ticket.
For this reason, accurate analysis and prediction and should be combined and corroborated with the other fields such as the Description.

The \textit{Description} field contains the full details of the issue.
When stakeholders want a detailed understanding of the issue (e.g., to be able to resolve it), they will go to the Description.
Using NLP to interpret the Description is essential and difficult, as descriptions come in many forms, from simple small paragraphs to multi-section deep dives.
Additionally, the form of information in Descriptions is not always natural language prose but can also include headers, code blocks, log outputs, steps-to-reproduce, screenshots and more.
In principle, all necessary information should be listed in the Description, but in reality, some information might be listed elsewhere in the issue, and some information might not be listed at all.
Additional information is often contained in the Comments.

\subsubsection{Metadata and Workflow.}
Issue Metadata covers fields that enhance the issue content and enable batch processing, retrieval, and workflow automation.
Metadata information is typically categorical data shared across the issues in a certain project or organisation (shared labels, environments, priority levels etc.), whereas Content information is unique.
The \textit{Labels} field contains tag-like information. For instance,  MCPE-31887\footnote{\url{https://bugs.mojang.com/browse/MCPE-31887}} has multiple labels referring to Minecraft items.
The \textit{Environment} field is label-like but focuses on hardware- or software-related information. 
This is often a text field, such as STL-1699\footnote{\url{https://jira.hyperledger.org/browse/STL-1699}} with the Environment set to ``On a MacBook Pro, using docker containers'', but it can be more technical, such as ``AWS
Marketplace 1.1.4 AMI'' on STL-1618.\footnote{\url{https://jira.hyperledger.org/browse/STL-1618}}
The \textit{VersionsAffected} field contains the software version(s) affected by the content described in the issue.
The \textit{VersionsFixed} field contains the software version(s) fixed by the content described in the issue.
For example QTBUG-117454\footnote{\url{https://bugreports.qt.io/browse/QTBUG-117454}}, its VersionsAffected is ``6.6.0 Beta4'' while its VersionsFixed is ``6.6.0 RC''.

Some metadata fields can be specific to the issue \textbf{workflow}, supporting and guiding issues in their lifecycle from creation to resolution.
The \textit{Status} field marks the current stage of the lifecycle an issue is in.
The \textit{Priority} field marks the importance of an issue in relation to other issues.
The \textit{Resolution} field marks the reason for closing an issue.
The \textit{CreatedDate} and \textit{ResolvedDate} mark the beginning and end of an issue lifecycle.
In Jira, the CreatedDate cannot change, and any changes to the ResolvedDate are not tracked in the evolutions.

Metadata and workflow fields contain short, possibly categorical natural language text, but sometimes with numbers as well.
These fields can either be learned and predicted based on historic data or they might enrich and extend the Summary and Description.
For example, a script could check for duplicate or inconsistent information in the Comments based on the current issue Labels.
Metadata should mostly be correct and consistent, but trackers often do not enforce validating their values.
Users might be able to, for example, change metadata without updating previous versions accordingly.

\subsubsection{Issue Tracker Structure.}
\label{sec:issue_structure}

The Structure fields provide structure to issue trackers, beyond simply annotating individual issues.
The \textit{IssueType} field describes the role this issue plays within the issue tracker: epic, story, bug, feature request, etc. (see Montgomery et al. ~\cite{Montgomery_MSR_2022} for a comprehensive breakdown of issue types).
The \textit{Project} and \textit{Components} fields provide a top-down structure to issue trackers that work roughly as mutually exclusive bins, e.g., every issue is under a Jira, Project, and Component(s).
The \textit{Parent} field contains a link to the parent issue that this issue is under.
The \textit{IssueLinks} field contains several links to other internal issues, resources, and even external issues in other issue repositories.

Many issue trackers offer the ability to link issues together via specialised link types.
These links relate issues to each other, allowing stakeholders to understand and capture the relationships between different issues, making it easier to find information, and structuring overall project knowledge~\cite{Sandusky_MSR_2004}.
For example, a reported issue might be part of a major bug, a bug might contribute to a specific requirement, or two issues might refer to the same feature request.
Issue linking can be considered \textit{horizontal traceability} as the links relate artefacts (issues) on the same level~\cite{Heck_JSEP_2014,Gotel_BOOK_2012}.
Issue links are listed either directly as an attribute on the corresponding issue (e.g. in JIRA and Bugzilla) or in the comments sections (e.g. in GitHub Issues).
There are different link types, such as Duplicate, Relate, Block, and Subtask \cite{Lueders_MSR_2022}, which serve different purposes and can be used to indicate various relationships between issues.

\subsubsection{Discussion and Community.}
A significant part of issue trackers and their data is the community of stakeholders and their involvement with the evolution of issues.
The \textit{Creator} field lists the person who opened the issue.
The \textit{Reporter} field contains the person who initially brought up the issue, which might be a different person than the creator.
The \textit{Assignee} field lists the person responsible for resolving the issue, which can change during the lifetime of the issue.
Other stakeholders might also register to simply \textit{Watch} the issue, to be notified when issue changes occur.
These fields can be combined with profiles of individuals or artefacts to being additional value to automated analysis.

One significant field is the \textit{Comments} field, which lists all the comments made on an issue.
For example, ``that could be fun to implement, although I won't have time for it for a while'' is a comment on QTBUG-76315.\footnote{\url{https://bugreports.qt.io/browse/QTBUG-76315}}
In this case, the developer is conveying interest, but also their relative time constraints (which is the likely reason for this issue still being open, despite being over 4 years old).
Another example is: ``I think you are right and since three months have passed, I can fix it for you; thank you for the issue and the suggested code changes'' on ZOOKEEPER-4703.\footnote{\url{https://issues.apache.org/jira/browse/ZOOKEEPER-4703}}
In most issue trackers, the comments are displayed in ascending chronological order after the Description.
Each comment tends to be just a few sentences long, but it can also be multiple paragraphs.
Comments might include questions, personal updates, or requests for information.

Some comments may contain relevant information not included in the Description.
For example, a comment could contain the log output of the error mentioned in the Description.
A comment could also include conflicting information relative to the Description or other fields.
For example, someone could comment that an open issue should have a status of ``Closed''.
Therefore, for a stakeholder to have up-to-date precise information about an issue, they might have to read all comments in addition to the Summary and Description.
When constructing machine representations of an issue (e.g. a vector embedding), it may benefit the algorithm to include information listed in the comments---in addition to the Summary and Description fields.

\subsubsection{Evolution Data.}
Some issue trackers like Jira also track the history of changes to the issue, on a per-field basis, with timestamps.
In Jira, each change is stored as a before and after of a field, including who changed it and when.
This data is described in Fig.~\ref{fig:database_erd} under ``item'', where the ``field'' evolution is described as a ``fromString'' and ``toString''.
Anything can change after the creation of an issue, including its core data and its metadata. Some fields, e.g. Status or Comments are more likely to change. 
Evolution data can be used to understand what happened to single issues, to predict values that are likely to change, or to understand a bigger picture in a project or organisation about what changes might occur in what context, why, and are they rather ``good or bad'' changes.

\section{Use Cases and Techniques for Mining Issue Trackers}

We present four major use cases for leveraging data mining and NLP to support stakeholders use their issue trackers.
For each use case, we introduce the underlying motivation, data analysis  algorithms applied, and general pitfalls.
We focus our discussion on Jira, while the use cases are likely relevant for any modern issue tracker.
Jira is one of the more popular issue trackers, and we have a large public dataset available to analyse~\cite{Montgomery_MSR_2022}.
We share a JupyterNotebook on Zenodo\footnote{\url{https://zenodo.org/doi/10.5281/zenodo.10779206}} with source code for three of the use cases.\footnote{The fourth use case already has a replication available at\\\url{https://github.com/RegenKordel/LYNX-BeyondDuplicates}}

\subsection{Issue Quality Analysis}
\label{sec:req_amb_det}

Requirements quality research aims to define, improve, and evaluate quality factors for natural language requirements.
These quality factors include ambiguity, completeness, consistency, correctness, and many more~\cite{Montgomery_REJ_2022}.
Research has produced techniques to automatically detect problems with such quality factors.
Issue trackers offer a well-suited environment to apply these techniques, as there is a diverse and plentiful set of requirements-related knowledge to analyse.

For this use case, we analyse User Story issues since they tend to be concise and atomic, making them easier to analyse than other requirements forms.
In the ``RedHat'' Jira Repository, there is a considerable number of user stories ({\mytilde}12,000 issues with the type ``Story'').
Not all 12,000 are strictly user stories, given the fluid nature of how issue trackers are used in practice.
Therefore, we filter out issues where the Description does not contain ``as a'' to remove issues that are not user stories.
This produces a table where each row is an issue, and the columns are the Jira issue ID and the Description.
User this data, we then detect the following ambiguities: subjective language and nominalisations.
We do this by applying different NLP techniques, including lexical lookups, regular expressions, part-of-speech tagging, parse trees, and hypernyms.
Generative Artificial Intelligence (GenAI) and Large Language Models (LLMs) have shown increasing promise as powerful tools for NLP applications.
These algorithms can likely be applied for language translation, grammar and spell checking, text classification, summarisation of long texts, and even full Description generation.
For this use case involving specific types of ambiguity, however, we will be focusing on traditional NLP techniques designed to target specific structural aspects of language.

\subsubsection{Subjective Language.}
There are multiple types of subjective language that are considered ambiguous.
Tjong et al.~\cite{Tjong_REFSQ_2013} and Gleich et al.~\cite{Gleich_REFSQ_2010} suggest algorithms to detect various types of ambiguous subjective language.

Tjong et al. addressed ``dangerous plurals'' and ``inside behaviour''~\cite{Tjong_REFSQ_2013}.
Dangerous plurals are subjective due to the lack of clarity regarding boundary conditions on words such as ``few'', ``little'', and ``many''.
Here is an example from our dataset: ``of course there's no need to provide every single detail of the proposed architecture in the design document'', where the use of ``every'' here is ambiguous, and left to stakeholders to interpret.
Inside behaviour is subjective due to the lack of specificity regarding the outside behaviour of functional elements, such as with the use of words like ``until'', ``during'', and ``after''.
Here is an example from our dataset: ``specific releases experience reliability issues during regular Motions'', where ``during'' here lacks a description of when these reliability issues start and stop.
Both types of subjective language can be detected using lexical lookups (searching for individual words).

Gleich et al. addressed ``unclear inclusion'' and ``passive voice''~\cite{Gleich_REFSQ_2010}.
Unclear inclusions are ambiguous due to the use of ``up to'' without the use of ``including'' or ``excluding''.
It is then unclear whether the described values are to be included or not in the definition of that functionality.
Here is an example from our dataset: ``the engine returns up to 10 fired events'', where it is unclear if the engine can return 10 events.
We can identify this ambiguity using the following regular expression:
{\footnotesize
\begin{verbatim}
    up\\sto\\s(?!.*including|excluding)
\end{verbatim}}


Gleich et al.~\cite{Gleich_REFSQ_2010} constructed a technique to detect ``passive voice'' with the use of part-of-speech (POS) tagging and regular expressions.
Passive voice leads to ambiguity where the actor of a verb is omitted, e.g., ``the gate was opened'' (but we do not know \textit{who} opened the gate).
To detect such an ambiguity, we need POS tags.
POS tagging is the process of assigning a word a tag that represents its role within the sentence, e.g., noun, verb, adjective, etc.
These tags are assigned based on the word itself and its context within the sentence.
Using POS tags allows us to search for more than just the literal words.
It allows us to search for classes of words and also words being used in specific ways (e.g., ``brush'' can be both a noun and a verb).
To detect passive voice, Gleich et al.~\cite{Gleich_REFSQ_2010} apply POS tagging to the requirement, searching for the verb ``to be'', followed by the past participle verb form (most often verbs ending with ``ed''), and no additional verbs between the two preceding parts.
You can see this three-part approach in the regular expression below (each part separated by a space).
{\footnotesize
\begin{verbatim}
    \\b\\w+?°V[^°]*°be (\\W[^°]+?°(?!VB.)[^°]*°[^ ]+?)* \\W\\w+?°VBN°\\w+
\end{verbatim}}

\subsubsection{Nominalisations.}

Nominalisation is the transformation of a verb phrase into a noun phrase, which reduces a complex process into a single noun word or phrase.
An alternative definition is given by Lapata where nominalisations are seen as ``a particular class of compound nouns whose head noun is derived from a verb and whose modifier is interpreted as an argument of this verb''~\cite{L02}.
Ambiguity can arise from nominalising a complex process since a different meaning (or meaning containing less information) is communicated~\cite{GR03}.
According to Goetz and Rupp~\cite{GR03}, nominalisations can manifest in requirements from distortions---i.e., the constant re-organisation of previously acquired knowledge.
For example, ``\textit{Powering down} the cab radio ...'' can be ambiguous because the complexity of ``Powering'' is hidden behind a noun and not adequately described.

To detect nominalisations, Bouraffa uses simple NLP methods, without the need of ad-hoc lookup tables~\cite{Bouraffa_TUHH_2019}.
They distinguish between two types of nominalisations: derived nominals and gerundive nominals.
Derived nominals originate from adverbs, adjectives, and verbs that are simplifying a complex process using additional \textit{derived} components, such as ``-tion'' or ``-ism.''
Gerundive nominals are verbs in the ``-ing'' form that are not being used as proper verbs in a sentence.
An example of a derived nominal and a gerundive nominal, is ``\textit{Powering down} the cab radio shall cause the \textit{disconnection} from the mobile network''.
``Powering'' is the gerundive nominal and ``disconnection'' is the derived nominal.
Both cases present complex processes that are simplified to single words.

Bouraffa identifies derived nominals using POS tags, special suffixes, and hypernyms~\cite{Bouraffa_TUHH_2019}.
A word suffix is a set of characters appended to the end of a word stem, such as ``-tion'', ``-ism'', or ``-ty''.
Hypernyms organize words in an \textit{is-a} relationship, creating a tree hierarchy that relates words to each other.
First, we retrieve nouns (i.e., POS tag ``NN'' and ``NNS'').
Then, we look for specific suffixes signalling a derived nominal.
Finally, since we are interested in the nominalisation of actions, we only retain words having ``EVENT'', ``PROCESS'', or ``ACT'' in their hypernym path.
These hypernyms indicate a noun that usually acts as a verb; hence, they have been nominalised.
Here is an example from our dataset: ``During \textit{default deployments} (without user-specific configuration) TLS termination is managed by the Route''.
``Default deployments'' is a nominalisation that hides a complex process and therefore introduces ambiguity, which is likely why the user had to add a qualifier in braces right after it.

Bouraffa identifies gerundive nominals using POS tags and dependency trees~\cite{Bouraffa_TUHH_2019}.
Dependency trees describe how words in a sentence are related to each other, for example, ``root'', ``compound'', and ``direct object.''
Using POS tags, we first retrieve terms tagged as ``VBG'' (i.e., a gerundive verb) and then check whether it is a gerundive nominal (nominalisation).
To this end, we parse the dependency tree and check if the VBG term is \textit{not} labelled as root verb of a sentence, auxiliary of a verb, adverbial modifier, compound, or clausal modifier of a noun.
These labels signify situations in which the gerundive verb is being used as an actual verb, and therefore is not a nominalisation. 
Here is an example from our dataset: ``Specific information about the \textit{persistence implementation} should not be in the CR attributes''.
The nominalisation is in ``persistence implementation'', which is a complex process that has reduced to a two-noun phrase, thereby hiding the details and creating ambiguity.

\subsubsection{Pitfalls and Takeaways.}

Issue trackers provide access to plenty of requirements data created by stakeholders with different backgrounds and expertises.
Before applying a mining or a machine learning technique to the core textual information, a manual checking and advanced preprocessing might be needed.
Issues labelled ``User Story'', for example, might not strictly follow User Story templates, and might contain surrounding context such as acceptance criteria.

Applying NLP techniques is then a matter of extracting the natural language text with the desired algorithms.
Some algorithms, such as Subjective Language algorithms, can produce false positives (requirements that are labelled as ambiguities but are actually not).
This is partly due to the nature of some algorithms using simple heuristics, partly due to the nature of ambiguity itself, and partly due to the heterogeneous data.
Research into ambiguity is still ongoing and even defining what is ambiguous, in which contexts, to whom, is not trivial.

\subsection{Evolution Analysis}

Evolution data is powerful as it facilitates viewing the data at any point in time and understanding what has changed (and possibly why).
This can be used as a tool for predictive modelling and grounded investigations, among other things.
Research into requirements evolution is limited mainly to change management, which is more about processes than data.
The concept of using requirements histories is not entirely novel, but it is rather understudied.

Requirements evolution data is quite rare since pre-finalised requirements are either not public, or non-systematically versioned.
Additionally, even when requirements do have evolution data, it is often in the form of requirements documents with multiple versions over time.
This can be challenging to unpack and utilise in research since tracing between requirements is frequently lost, and it is not trivial to write automatic extraction scripts.  
Thus, issue trackers offer a unique source of requirements evolution data since issues and their evolutions are well-structured.
Additionally, issues are designed to be atomic units of information, so individual requirements tend to be in individual issues.

Analysing the evolution depends on the particular evolving field being analysed. 
In particular, the evolution of issue metadata, issue structure, and issue content might require different techniques and might lead to different insights.
Concerning metadata and structure, there is a large body of work from the last decade.
The overall approach is usually the same: predict the value of a certain issue field based on learning from past data.
For instance, we could train a machine learning classifier on past issue data to predict which priority an issue is likely to have, who should be the assignee~\cite{Stanik:ICSME:2018}, or what link to other issue is should have~\cite{Lueders_REJ_2023}.
Most of these works use the last snapshot of previous issues to predict the field of a new issue (or just randomly split issues into training and testing sets without considering the time).
The \textit{evolutions} of the issues in the training set are typically ignored.
This might, however, bear useful information about false positive predictions or of what fields are likely to change. 

Applying evolution analysis on core textual data of the issues might lead to insight about when an issue (or a requirement) changes and why.
One such a reason could be that the text was ambiguous in the original text, contextual information was missing, or the language is ``too emotional''. 
Sentiment analysis rates the affective nature of the text as either positive, negative, or neutral.
In other words, is the writer of the text conveying positive, negative, or neutral emotion through their words.
Certain sentiment analysis tools extract more information than that, including specific emotions such as happiness, sadness, or anger.
However, these tools often require more input data for a confident prediction and the results are often less accurate than sentiment polarity alone.
One objective is to investigate sentiment across evolutions and uncover the dynamics of issue descriptions through time.
In particular, we are interested to see if edits make descriptions rather neutral in sentiment.
Issue descriptions should be strictly informational, so perhaps over time the sentiment trends to zero.

Any requirement type from any Jira repository can fit to this analysis.
We thus sample issues from all available repositories.
We first gather 10,000 issues per repository, randomly, and then perform several cleaning steps to remove defects such as missing and impossible data (e.g., closed date before open date).
There are then 64,840 issues left, with most Jira repos containing {\mytilde}5--7k issues, and some smaller repos such as SecondLife and Sonatype contributing {\mytilde}500 issues.
We first want to consider how many evolutions each issue has.
At a minimum, each issue should have at least one evolution where the description changes.
Finally, we filter the data down to just the Description field, since we are only analysing that text.
Our final dataset is a table with 35k rows, where each row is an evolution to an issue's Description field.

Now that we have a table with a row for every evolution to Descriptions, we extract the sentiment score and track that over time as an issue evolves.
For simplicity, we compare the sentiment of the first version of the description to the last version.
We extract the sentiment score using SpaCy spacytextblob package.\footnote{\url{https://spacy.io/universe/project/spacy-textblob}}
Finally, we calculate the description sentiment ``trend'' and store that alongside the descriptions themselves, to allow for manual analysis.

Here are two examples of issue descriptions at the start and end, and the sentiment trend between them.
The description at creation was: ``Chef has been identified as a suitable option to create the automatic deployment mechanism. The architecture of the mechanism needs to be specified, and then implement in Chef''.
The final description simply had ``suitable'' removed, leading to a 0.55 drop in sentiment, which means the sentiment became more negative.
This makes sense, since ``suitable'' is a positive word, and removing it made the description more technical and without pleasantry.
Another issue description started as: ``The Wss4jSecurityInterceptor has no X509 Binary security token support yet. It would be great if we could add it.''
The description at the end was ``Document the X509 Binary security token support for Wss4j.'', which resulted in a sentiment drop of 0.8.
This appears, again, to be a matter of removing positive words such as ``great'' in favour of a more literal and to-the-point description.

\subsubsection{Pitfalls and Takeaways.}
Issue trackers offer a specifically powerful perspective on requirements with their full evolutionary history.
We have showcased one such analysis of this data, but there are many more opportunities.
For sentiment analysis, it appears that the sentiment model is very sensitive to non-textual elements, such as URLs.
For this reason, it is important to filter the data appropriately and do many manual checks of the results before coming to conclusions.
Additionally, issue trackers have many unexplainable data points that can break even robust analysis tools.
This data includes strings that are too long, likely due to log dumps or copy-paste content from software output.

\subsection{Discussion Analysis}
Requirements engineering is a collaborative process involving multiple stakeholders who interact to clarify, prioritize, and implement requirements.  
Issue trackers offer a central place for analysing and assisting this collaboration~\cite{Bertram_CSCW_2010}.
It is common that each issue has a running list of comments just below the description.
Even in more restrictive issue trackers where only certain people can create or edit issues, the ability to comment is often left open to the public.

There are various objectives for such analysis. First, discussion analysis can monitor the overall activity intensity, as some issues tend to receive more comments than others. Here similar sentiment analysis can be applied as to the issue description evolution, e.g. to detect and possibly steer escalations. 
Second, comments might include different types of information which can be useful for different tasks and different stakeholders, such as steps to reproduce, workarounds, background knowledge and rationale why a certain feature is needed etc. Discussion analysis can aim at classifying or retrieving such specific information to support an easier access to the information~\cite{Arya_ICSE_2019}. 
Third, information in the comments that should be propagated to the issue fields can be extracted, and the fields updated.
For example, situations where someone has commented that the status should be updated, but it has yet to be.
To demonstrate such analysis, we use the same dataset that we generated for the last use case: a sample of issues from all Jira repositories in the dataset.
However, we do not filter the issues or evolutions on a particular condition.

Our objective is to find cases where the comments or description mention attributes that should be changed in their respective fields.
This task is quite difficult with high precision, so we aim for high recall and do some fast manual analysis to find what we are really looking for.
The first step towards our objective is to gather a list of all field names to search for in the comments and descriptions.
This is trivial: we gather a unique set of all ``field'' names from our table leading to 26 unique fields.
We can already search for these field names in comments and descriptions, which are likely candidates for related discussions.
However, to further reduce the dataset and increase precision (while maintaining high recall) we require that the comment or description mentions one of the values for this field (we call these states).
For example, ``I think the \textit{status} is wrong'' would not be enough, it would have to say something like ``I think the \textit{status} should be set to \textit{closed}''.
To find all possible states for a given field, we gather a unique set of all values that field has ever been set to.
To keep the lists manageable and realistic, we limit each list to all states that field has ever been, \textit{in the Jira it is in}.
The result is a multi-level dictionary item where for a given Jira and field, all states are listed.

We then mine all comments and description evolutions, searching for cases where both a field name has been mentioned, \textit{and} at least one of the field states that it could be in.
Our analysis finds many results; however, as previously mentioned, we now need to manually filter through them to identify situations where there is really an explicit suggestion for a field to have a different value.
Here is an example of a false positive: ``Sorry, this is not a \textit{Major} \textit{Priority}, so I cannot demote''.
They happen to use the words Priority (field name) and Major (possible field state), but they are not recommending that the priority field be changed to major.
There are also many close but still false positives where a comment retroactively comments that a field has been changed, e.g., ``Changed Priority to Major''.
One example of a true positive is: ``I do not know why the Priority is changed to P2-High'', which is discussion point where the author thinks the priority should be changed to something else.
Here is a true positive we aimed at identifying: ``We need to fix this issue ASAP. [...] Please mark this issue as High Priority''.
Here is another true positive: ``That's not a Low Priority ticket because our workflow is completely blocked right now''.
With such a simple lookup, we can filter down the set to a reasonably sized set with pre-conditions we assume are necessary  to find the right data.

\subsubsection{Pitfalls and Takeaways.}
Analysing issue discussions does not require advanced NLP techniques. 
Already, word matching to find terms reflecting issue statuses, priorities, or related issues can lead to identifying relevant comments with a satisfying precision.
Certainly, more advance information retrieval techniques, text classification, and semantic reasoning could improve the output.

\subsection{Link and Structure Analysis}

Link analysis serves as an approach for gaining insights into the relationships between two issues. When applied recursively, more coarse-grained structure analysis can be achieved, e.g. to identify issue sub-graphs in a project denoting workflow, organisational or architectural dependency \cite{Lueders_MSR_2022}.

Issue trackers enable users to connect the various pieces of information they store in issues.
Therefore, issue trackers are good sources for conducting link analysis due to their comprehensive tracking of both issues and the links between them.
One of the most prominent research problems is duplicate detection~\cite{Deshmukh_ICSME_2017,Anvik_OOPSLA_2005,He_ICPC_2020,Wang_ICSE_2008}, which involves determining whether two issues essentially describe the same problem or feature. 
This task can be approached using NLP techniques that assess textual or semantic similarity between issues, analyse differences in text length, or consider the combined length of text.
However, beyond duplicate detection, issue trackers encompass various link types providing an extensive landscape for exploration.
This type of analysis is particularly beneficial for users navigating large issue trackers with substantial backlogs, as it aids in the discovery of relevant issues amid the vast array of available data.

One primary objective in conducting this analysis is to gain insights into the similarities and distinctions among linked texts based on their types.
For the purposes of the demonstration, we focus  on issues that are created within the year of 2021 and that have established links to other issues.
Furthermore, we focus on repositories with include prevalent links and multiple link types.
The first step for issue link analysis is to represent the issues as vectors which can compared in a particular space: a technique that is called embedding.  
Embeddings take words or sentences and try to embed them into a vector space. 
Words or sentences with similar meanings should be close to each other and words or sentences with very dissimilar meanings should be far away from each other.
The textual similarity between pairs of linked issue texts can be measured using the cosine similarity: that is the angle between two vectors divided by the product of their lengths. This results in a similarity score from $[-1,1]$.

\begin{figure}
    \centering
    \includegraphics[width=1\textwidth]{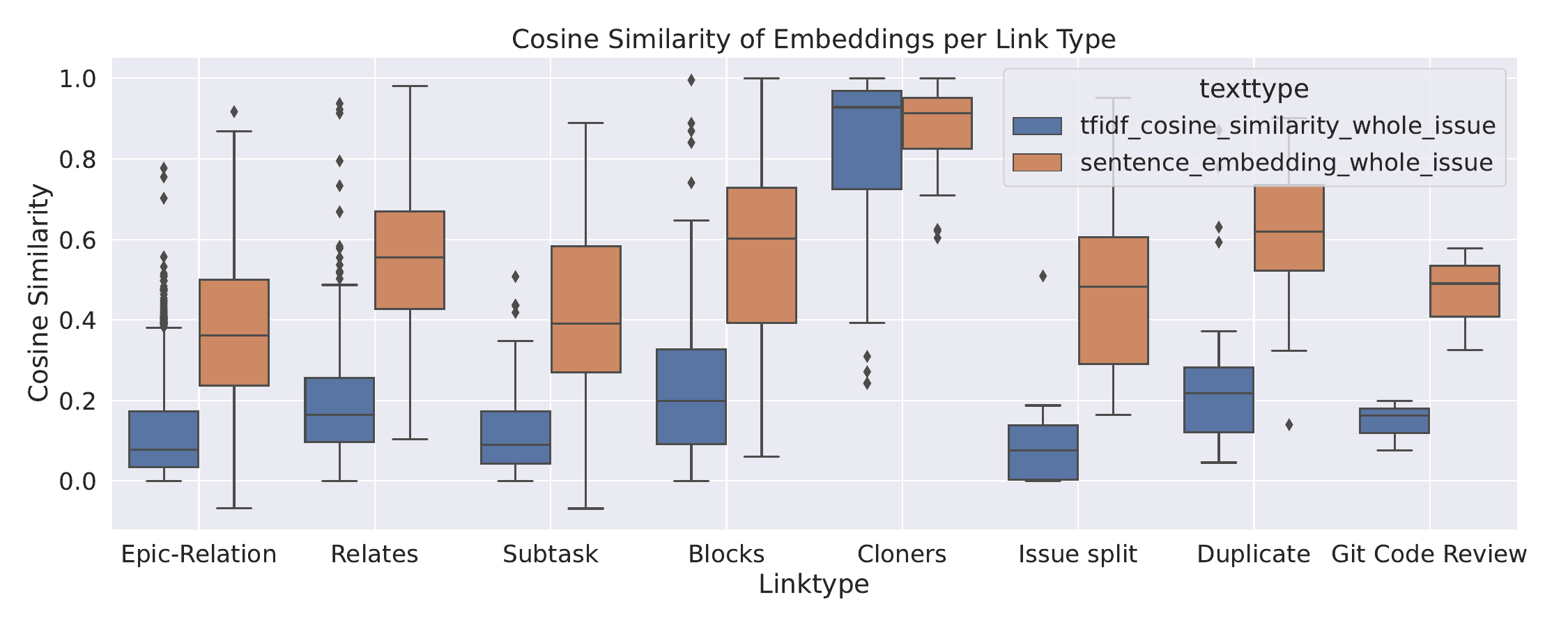} 
    \caption{Link analysis with TF-IDF and BERT embeddings and their cosine similarity of issue texts of linked issues per link type.}
    \label{fig:linkanalysis}
\end{figure}

We can employ two embedding techniques: TF-IDF and BERT. 
We embed the text content, encompassing both the issue titles and their respective descriptions. 
We can compare the cosine similarity scores derived from the TF-IDF embeddings with those produced by the BERT embeddings, with a specific focus on various link types identified within the Hyperledger repository.
Figure~\ref{fig:linkanalysis} shows the distribution of the cosine similarities per link type.

In our analysis, we have made several observations. 
First, we noticed a consistent trend: the cosine similarity of TF-IDF embeddings tends to be consistently lower than that of BERT embeddings. 
This can be attributed to the inherent limitations of TF-IDF in capturing nuanced semantic meanings and synonyms when compared to BERT. 
BERT, being a contextual language model, excels in comprehending contextual and semantic information within text. 
Consequently, BERT embeddings often yield higher cosine similarity scores, indicating a more thorough grasp of the underlying semantics of the text.

Second, within our dataset, we identified a distinct category known as the \textit{Cloners}. 
These issues exhibit the highest similarity, regardless of whether TF-IDF or BERT embeddings are used. 
This elevated similarity is largely attributed to the nature of \textit{Cloners}. 
These issues are typically created with Jira's clone feature, allowing users to generate new issues based on existing ones while retaining the flexibility to modify specific attributes. 
Consequently, \textit{Cloners} inherit a substantial portion of content from the original issue, particularly in terms of TF-IDF vectorisation, which contributes to their high TF-IDF similarity.

Third, beyond the \textit{Cloners}, we found that other link types display relatively similar cosine similarity scores, regardless of the embedding technique employed. 
For TF-IDF, the scores range from 0.12 to 0.25, while for BERT, they span from 0.36 to 0.60. 
Interestingly, the \textit{Duplicate} link type does not significantly differ in its cosine similarity compared to other link types. 
This observation raises questions about the effectiveness of models solely relying on text similarity in distinguishing \textit{Duplicate} links from other types, as noted by L{\"u}ders et al.~\cite{Lueders_MSR_2022}.

General typed link detection remains an active research area. 
Previous studies, such as those by L{\"u}ders et al.~\cite{Lueders_REJ_2023} (macro F1-scores ranging from 0.41 to 0.88 and weighted F1-scores of 0.64 to 0.95) and Nicholson et al.~\cite{Nicholson_AIRE_2020,Nicholson_AIRE_2021} (weighted F1-scores from 0.56 to 0.70) showcasing the ongoing challenges. 
L{\"u}ders et al.~\cite{Lueders_REJ_2023} used the same Jira dataset. Their replication package can be used to demonstrate link classification. 
Borg et al.~\cite{Borg_ECSMR_2013} categorised explicit links in two issue-tracking systems into four categories: ``Related'', ``Duplicate'', ``Clone'', and ``Miscellaneous'', highlighting that ``Clone'' links represent a stronger form of ``Duplicate'' links due to their issues sharing identical textual content. 
Li et al.~\cite{Li_APSEC_2018} delved into issue linking practices on GitHub, categorising link types into six categories, including ``Dependent'', ``Duplicate'', ``Relevant'', ``Referenced'', ``Fixed'', and ``Enhanced''.
Their findings emphasised the need for automated classification, where ``Referenced'' links often referred to historical comments with essential knowledge, and ``Duplicate'' links were frequently marked within the same day. 
Finally, Tomova et al.~\cite{Tomova_ICSE_2018} investigated seven open-source systems, revealing that the rationale behind selecting a specific link type is not always obvious. Additionally, they noted that \textit{Clone}~links are indicative of textual similarity, while issues connected through a \textit{Relate}~link exhibit varying degrees of textual similarity, often necessitating additional contextual information for accurate identification.

\section{Discussion and Summary}
Issue trackers have emerged as a new key place where requirements knowledge is crafted, delivered, and maintained---along with bugs, tasks and other system and project knowledge. 
What makes issue repository even more interesting is that issues usually are connected to source code repository and other resources. 
Therefore, mining issue trackers as a rich and integrated knowledge base for software repositories bears a large potential to assist various stakeholders including analysis, product owners, developers, and users.
Capturing, managing and sharing requirements knowledge has been an ongoing challenge and area of research for decades~\cite{maalej2013managing}.
With issue trackers, requirements knowledge is directly connected in complex linked graphs, traceable directly to code, and explained through rationale links back to feature requests.
The capability of these systems is not just contained to what can be extracted from them, issue trackers are also a rich platform for embedded recommender systems~\cite{Herzig_BOOK_2014} in the form of browser extensions and tool plugins.

Issue trackers continue to fill up with information over time, which inevitably causes information overload and makes navigating these systems challenging.
Managing and extracting insights from the vast amount of heterogeneous data can be overwhelming.
Moreover, the quality of textual information within issue trackers often varies widely.
Inconsistent information, low-quality text, and too large ``knowledge graphs'' can hinder the effectiveness of mining efforts and using the data.
We demonstrated how properly leveraging NLP techniques, even as simple as term look-ups, can help unpacking, utilising, and even correcting the information stored in issue trackers.

NLP can help in addressing many RE-related use cases, but the essential component of this lies in the linked meta-data.
Traditional NLP and modern machine learning techniques, such embeddings, semantic similarity, and language generation offer effective methods to analyse, extract, and correct information from textual data in issue trackers.
Alongside the textual data, issue trackers offer a rich set of meta-data that can enhance solutions beyond just the insights available in the text.
Holistic approaches combining NLP with the meta-data can offer much more value in approaching more complex problems.


%
%
%
\bibliographystyle{splncs04}
\bibliography{main}
\end{document}